\documentclass[aps,twocolumn,epsfig,graphics,showpacs,floatfix,mathbbm]{revtex4}
\usepackage{amsmath,epsfig}
\usepackage{psfrag,times}
\begin{document}
\title{Quantum dense coding scheme via cavity decay}
\author{Zheng-Yuan Xue\footnote{E-mail: zyxue@ahu.edu.cn}, You-Min Yi and Zhuo-Liang
Cao\footnote{E-mail: zlcao@ahu.edu.cn}}

\affiliation {School of Physics and Material Science, Anhui
University, Hefei, 230039, China}
\begin{abstract}
We investigate a secure scheme for implementing quantum dense coding
via cavity decay and liner optics devices. Our scheme combines two
distinct advantages: atomic qubit sevres as stationary bit and
photonic qubit as flying bit, thus it is suitable for long distant
quantum communication.
\end{abstract}


\maketitle

\section{Introduction}
Quantum entanglement, a fundamental feature of many-body quantum
mechanical systems, was regarded as a key resource for many tasks in
quantum information processing \cite{dense,teleport,secret}. Quantum
dense coding (QDC) is a process to send two classical bits (cbits)
of information from a sender (Alice) to a remote receiver (Bob) by
sending only a single qubit. It works in the following way.
Initially, Alice and Bob shared a maximally entangled state. The
first step is an encoding process where Alice performs one of the
four local operations on her qubit. Then she sends the qubit to Bob.
The last step is a decoding process. After Bob received the qubit,
he can discriminate the local operation of Alice by using only local
operations, \emph{i.e}., Bell state measurement in the work of
Bennett \textit{et al}. \cite{dense}. Later, due to its predominate
importance in quantum communication, QDC attracts many public
attentions. Barenco and Ekert \cite{4} first addressed the question
of QDC with partially entangled state, and they focused on
deterministic QDC and considered the classical capacity of it.
Conversely, one can also consider the case of QDC by partial
entangled state with maximal classical capacity (2 cbits per qubit)
in a non-deterministic or probabilistic way \cite{5}. On the other
hand, QDC has been experimentally demonstrated using optical systems
\cite{6}, nuclear magnetic resonance (NMR) techniques \cite{7} and
trapped ions systems \cite{ion}. But, in the conventional process of
QDC the receiver can always successfully cheat Alice if he want and
use the information willingly. Now, the question arises, is there
any secure QDC scheme exist? Fortunately, we note that quanutm
secret sharing (QSS) \cite{secret} is likely to help in protecting
secret information. Here, in analogy with QSS, we term \emph{secure
QDC} as a process securely distributing information via QDC among
many parties in a way only when they cooperate can they read the
distributed secret information.

In the realm of atom, cavity quantum electrodynamics (QED)
techniques has been proved to be a promising candidate for the
physical realization of quantum information processing \cite{8}. The
cavities usually act as memories, thus the decoherence of the cavity
field becomes one of the main obstacles for the implementation of
quantum information in cavity QED. Recently, Zheng and Guo proposed
a novel scheme \cite{9}, which greatly prolong the efficient
decoherence time of the cavity with a virtually excited nonresonant
cavity. Osnaghi \textit{et al}. \cite{10} had experimentally
implemented the scheme using two Rydberg atoms crossing a
nonresonant cavity. Following these progresses, schemes for
implementing QDC are also proposed \cite{11}, where atomic qubit is
used as both stationary and flying bit. But, it is well known that
atomic qubits are only ideal stationary qubits, not suitable for
long distance transmission. Thus the realization of long distance
quantum communication with atomic qubits serving as flying qubits as
suggested in \cite{11} is an difficult experimental challenge.
Meanwhile, spontaneous and detected decay are unavoidable in
practical quantum information processing in cavity QED system
\cite{12}. However, it is shown recently that the detection (or the
non detection) of decays can be used to entangle the states of
distinct atoms \cite{13}. Furthermore, it can be used for quantum
communication protocols such as QT \cite{14, 15}, as Photonic qubits
are perfect candidate for flying qubits.

Here, we investigate a physical scheme for implementing QDC via
cavity decay and liner optics devices. The motivation of this work
is twofold: investigating physical QDC scheme via cavity QED
technology and solving the problem of security in practical QDC
schemes. Our scheme combines two distinct advantages: atomic qubit
sevres as stationary qubit and photonic qubit as flying qubit, thus
it is suitable for long distance quantum communication. We
investigate the scheme in a way similar to QSS \cite{secret} via a
tripartite entangled GHZ state, thus it is also a secure one. We
firstly consider the atom-cavity interaction in section 2, then
provide our scheme in section 3. Section 4 is some discussions about
our scheme and summary of our paper.

\section{Atom-cavity interaction}
\begin{figure}[tbp]
\includegraphics[scale=0.35, angle=90]{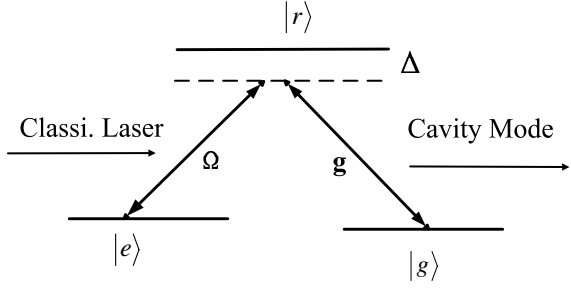}
\caption{Atom level structure in our scheme. The
$|e\rangle\rightarrow|r\rangle$ transition is driven by a classical
laser pulse with coupling $\Omega$, and the
$|r\rangle\rightarrow|g\rangle$ transition is driven by the
quantized cavity mode with coupling $g$. Both the classical laser
pulse and the cavity mode are detuned from their respective
transition frequencies by a same amount $\Delta$.} \label{fig1}
\end{figure}

Here and afterwards, each of the atoms has a three-level structure,
which has two ground states $|g\rangle$,$|e\rangle$ (\emph{e.g.}
hyperfine ground states) and an excited state $|r\rangle$ as
depicted in Fig.(\ref{fig1}).
It is an adiabatic evolution for the $|e\rangle\rightarrow|r\rangle$
transition, which is driven by a classical laser pulse with coupling
coefficient $\Omega$. The $|r\rangle\rightarrow|g\rangle$ transition
is driven by the quantized cavity mode with coupling coefficient
$g$. Both the classical laser pulse and the cavity mode are detuned
from their respective transition frequencies by the same amount
$\Delta$. Assuming the atom is trapped in a specific position in the
cavity, and the coupling coefficients $\Omega$ and $g$ are constant
during the interaction. In the case of $\Omega g/\Delta ^2 < < 1$,
the upper level $|r\rangle$ can be decoupled from the evolution.
When $\Delta>>\gamma$ the spontaneous decay rate $\gamma$ from
$|r\rangle$ level can be neglected \cite{14}. Suppose $\Omega=g$,
the effective Hamiltonian, in the interaction picture, is
\begin{equation}
\label{h} H_{e}=i\delta(a|e\rangle\langle g |-a^+|g\rangle \langle
e|)-ika^+ a,
\end{equation}
where $\delta=g\Omega/\Delta$, $a$ and $a^+$ are the annihilation
and creation operators of the cavity mode, $k$ is the photon decay
rate from the cavity. The time evolution of the interaction under
the Hamiltonian in Eq. (\ref{h}) are
\begin{subequations}
\label{e}
\begin{equation}
|g\rangle|0\rangle\rightarrow|g\rangle|0\rangle,
\end{equation}
\begin{equation}
\label{e2}
|e\rangle|0\rangle\rightarrow(\alpha|e\rangle|0\rangle+\beta|g\rangle|1\rangle).
\end{equation}
\end{subequations}
where we have discard the phase factor, which can be removed by a
simple rotate operation. The coefficients in Eq. (\ref{e2}) are
$$\alpha=e^{-\frac{1}{2}kt}(\cos{\frac{1}{2}}\Omega_{k}t+\frac{\Omega}{\Omega_{k}}\sin{\frac{1}{2}\Omega_{k}t}),$$
$$\beta=-\frac{2\Omega}{\Omega_{k}}e^{-\frac{1}{2}kt}\sin{\frac{1}{2}}\Omega_{k}t$$
with $\Omega _k = \sqrt {4\delta^{2} - k^{2}}$.

\section{Secure QDC with tripartite GHZ state}
Suppose Alice wants to send secret information to a distant receiver
Bob. As she does not know whether he is honest or not, she makes the
information shared by two users, \textit{i.e}., Bob and Charlie. If
and only if they collaborate, one of the users can read the
information, furthermore, individual users could not do any damage
to the process. The sender can probabilistically transmit two cbits
of information by sending only one qubit to the two receivers. By
collaboration, one of them could obtain the exact information,
furthermore, any attempt to obtain the secret information without
cooperation cannot succeed in a deterministic way. Assume the three
parties, \textit{i.e}. Alice, Bob and Charlie, initially share a
tripartite entanglement, which has been prepared \cite{guo} in the
GHZ type entangled state
\begin{equation}
\label{GHZ}
|\psi\rangle_{1,2,3}=\frac{1}{\sqrt{2}}(|eee\rangle+|ggg\rangle)_{1,2,3},
\end{equation}
where  $|e\rangle$ and $|g\rangle$  are the excited and ground
states of the atoms, respectively. Atoms 1, 2 and 3 belong to Alice,
Bob and Charlie, respectively.

Step 1. Alice decides to select one of the following two possible
choices. With probability $p$ Alice selects the first choice of
security checking, which aims to check the security of quantum
channel, and then the procedure continues to Step 2. Otherwise,
Alice can also move to the information encoding step with
probability $(1-p)$, the aim of which is to encode  and implement
the QDC procedure. In this case, the procedure goes to Step 3.

Step 2. Security checking. Hillery \textit{et al}. \cite{secret}
show that tripartite entangled GHZ state is sufficient to detect a
potential eavesdropper in the channel. In other words, the
eavesdropper could not succeed in a deterministic way during the QDC
procedure.

Step 3. Information encoding. Alice performs one of the four local
operations $\{\textit{I}, \sigma^{x},i\sigma^{y}, \sigma^{z}\}$  on
her atom. These operations denote 2 cbits information, and will
transform the state (\ref{GHZ}) to
\begin{subequations}
\label{GHZa}
\begin{equation}
 |\psi\rangle_{1,2,3}=\frac{1}{\sqrt{2}}(|eee\rangle+|ggg\rangle)_{1,2,3},
\end{equation}
\begin{equation}
 |\psi\rangle_{1,2,3}=\frac{1}{\sqrt{2}}(|gee\rangle+|egg\rangle)_{1,2,3},
\end{equation}
\begin{equation}
 |\psi\rangle_{1,2,3}=\frac{1}{\sqrt{2}}(|gee\rangle-|egg\rangle)_{1,2,3},
\end{equation}
\begin{equation}
|\psi\rangle_{1,2,3}=\frac{1}{\sqrt{2}}(|eee\rangle-|ggg\rangle)_{1,2,3}.
\end{equation}
\end{subequations}
Now the information is encoded into the pure entangled state, which
is shared among the three parties, and the encoding of the two cbits
information is completed.

\begin{figure}[tbp]
\includegraphics[scale=0.28,angle=90]{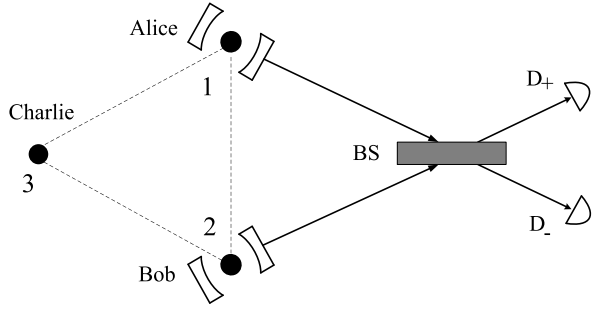}
\caption{The setup is adapted to implement the secure QDC scheme.
Atoms 1, 2 and 3 are initially prepared in tripartite entangled
state. Label the cavity of Alice and Bob as $A$ and $B$,
respectively, and they were both initially prepared in the vacuum
state. The 50/50 beam splitter (BS) and single-photon detectors
$\mbox{D}_\pm$ are located in Bob's side.} \label{fig2}
\end{figure}

Step 4. Information extracting. Alice applies a classical laser
pulse on atom to switch on the effective Hamiltonian $H_{e} $ of
atom 1 and cavity A, and then lead the photonic qubit flying to one
of the two receivers (\textit{i.e}., Bob). We will latter discuss
which is the party Alice sends her qubit to is not arbitrary. After
a party receives the qubit, he will have a higher probability of
successful cheat compared with the one who have not in the QDC
procedure. So, Alice would send her atom to the party, which is less
likely to cheat. We will discuss this latter in detail. Assume Bob
was selected to receive Alice's photonic qubit, and both cavities
are initially prepared in vacuum state. The setup of our scheme is
shown in figure (\ref{fig2}).

Meanwhile, Bob also switch on the effective Hamiltonian of atom 2
and cavity B. An interaction time of $\tan \frac{\Omega _{k} t}{2} =
-\frac{\Omega _{k}}{k}$ for both systerm leads the state
(\ref{GHZa}) to
\begin{subequations}
\label{GHZb}
\begin{equation}
 |\psi\rangle_{A,B,3}=\frac{1}{\sqrt{2}}(\beta^{2}|11\rangle_{A,B}|e\rangle_{3}+|00\rangle_{A,B}|g\rangle_{3}),
\end{equation}
\begin{equation}
 |\psi\rangle_{A,B,3}=\frac{\beta}{\sqrt{2}}(|01\rangle_{A,B}|e\rangle_{3}+|10\rangle_{A,B}|g\rangle_{3}),
\end{equation}
\begin{equation}
  |\psi\rangle_{A,B,3}=\frac{\beta}{\sqrt{2}}(|01\rangle_{A,B}|e\rangle_{3}-|10\rangle_{A,B}|g\rangle_{3}),
\end{equation}
\begin{equation}
|\psi\rangle_{A,B,3}=\frac{1}{\sqrt{2}}(\beta^{2}|11\rangle_{A,B}|e\rangle_{3}-|00\rangle_{A,B}|g\rangle_{3}).
\end{equation}
\end{subequations}
where the subscript of $A$ and $B$ corresponds to the cavity of
Alice and Bob, respectively. We have omitted the state of atoms 1
and 2 in Eq. (\ref{GHZb}) as they disentangled with the  subsystem
of (A,B,3). After the controlled interaction, which can be
controlled easily by a velocity selector \cite{8}, the states of
atoms 1 and 2 are both in their ground states.

Charlie let his atom crosses a classical field tuned to the
transitions $|g\rangle\leftrightarrow|e\rangle$. Choose the
amplitudes and phases of the classical fields appropriately so that
atom 3 undergoes the following transitions
\begin{subequations}
\begin{equation}
 |e\rangle\rightarrow \frac{1}{\sqrt{2}}(|e\rangle_{3}+|g\rangle_{3}),
\end{equation}
\begin{equation}
|g\rangle\rightarrow
\frac{1}{\sqrt{2}}(|e\rangle_{3}-|g\rangle_{3}),
\end{equation}
\end{subequations}
which leads the state in Eq. (\ref{GHZb}) into
\begin{subequations}
\label{GHZc}
\begin{equation}
 |\psi\rangle_{A,B,3}={\sqrt\frac{\beta^{4}+1}{2}}(|\phi^{+}\rangle_{A,B}|e\rangle_{3}+|\phi^{-}\rangle_{A,B}|g\rangle_{3}),
\end{equation}
\begin{equation}
 |\psi\rangle_{A,B,3}=\frac{\beta}{\sqrt{2}}(|\psi^{+}\rangle_{A,B}|e\rangle_{3}+|\psi^{-}\rangle_{A,B}|g\rangle_{3}),
\end{equation}
\begin{equation}
  |\psi\rangle_{A,B,3}=\frac{\beta}{\sqrt{2}}(|\psi^{-}\rangle_{A,B}|e\rangle_{3}+|\psi^{+}\rangle_{A,B}|g\rangle_{3}),
\end{equation}
\begin{equation}
|\psi\rangle_{A,B,3}={\sqrt\frac{\beta^{4}+1}{2}}(|\phi^{-}\rangle_{A,B}|e\rangle_{3}+|\phi^{+}\rangle_{A,B}|g\rangle_{3}).
\end{equation}
\end{subequations}
where
\begin{subequations}
\label{ab}
\begin{equation}
\label{ab1}
 |\psi^{\pm}\rangle_{A,B}=\frac{1}{\sqrt{2}}(|01\rangle_{A,B}\pm|10\rangle_{A,B},
\end{equation}
\begin{equation}
\label{ab2}
 |\phi^{\pm}\rangle_{A,B}=\frac{1}{\sqrt{\beta^{4}+1}}(\beta^{2}|11\rangle_{A,B}\pm|00\rangle_{A,B}.
\end{equation}
\end{subequations}
Obviously, one can see that there is an explicit correspondence
between Alice's operation and the measurements results of the two
receivers, which means that if they cooperate, both of them can read
the information. But if they do not choose to cooperate, neither of
the two users could obtain the information by local operation in a
deterministic manner. In this way, we complete the procedure of
secret extraction, and the above procedures from step 1 to step 4
constitute a complete process of secure QDC. One can repeat the
above procedures until all the information was sent.

Then the only task is to discriminate the four state in Eq.
(\ref{ab}). We note that the two states in Eq. (\ref{ab1}) can be
easily discriminated \cite{16} from Eq. (\ref{ab}), and the
implementation is shown in figure (\ref{fig1}). The discrimination
of Eq. (\ref{ab2}) from Eq. (\ref{ab}) requires photon number
discrimination detector, which is a technique still under extensive
exploration. So, our scheme is a probabilistic one.

Next, we consider the discrimination of Eq. (\ref{ab1}) from Eq.
(\ref{ab}). They are discriminated by the different clicks of the
two single-photon detectors as shown in figure (\ref{fig1}). Before
one of the two detectors clicks, the state (\ref{ab1}) will evolve
to \cite{plenio}
\begin{subequations}
\begin{equation}
\label{ab11}
 |\psi(t)^{+}\rangle_{A,B}=\frac{1}{\sqrt{2}}e^{-kt}(|01\rangle_{A,B}+|10\rangle_{A,B}.
\end{equation}
\begin{equation}
\label{ab12}
 |\psi(t)^{-}\rangle_{A,B}=\frac{1}{\sqrt{2}}e^{-kt}(|01\rangle_{A,B}-|10\rangle_{A,B}.
\end{equation}
\end{subequations}
While one of the detectors $\mbox{D}_\pm$ clicks, it corresponds to
the action of the jump operators $1/\sqrt2(a_{A} \pm a_{B})$ on the
joint state $|\psi(t)^{\pm}\rangle$. The click of $\mbox{D}_{+}$ and
$\mbox{D}_{-}$ correspond to the states of (\ref{ab11}) and
(\ref{ab12}), respectively. Based on the above analysis, the total
probability of successfully discriminate the four states in Eq.
(\ref{ab}) is $\beta^{2}e^{-2kt}$, in other words, our scheme
succeed with a probability of $\beta^{2}e^{-2kt}$.

\section{Discussions and summary}
Now, let's turn to the case if they do not choose to cooperate with
each other. Without the cooperation of Charlie, Bob knows Alice's
operation belongs to one of $\{\sigma^{x},i\sigma^{y}\}$  with unit
probability when the detector clicks. But he cannot further
discriminate which one Alice's operation is. In other words, without
the cooperation of Bob, Charlie knows nothing further about Alice's
operation. If Charlie lies to Bob, Bob also has a probability of
$\frac{1}{2}$  to get the correct information, so the successful
cheat probability of Charlie is $\frac{1}{2}$. Conversely, Charlie
only has a probability of $\frac{1}{4}$  to get the correct
information, so the successful probability of Bob is  $\frac{3}{4}$.
This is the point that we have mentioned in the beginning of Step 4,
that is, he who received Alice's atom has a higher probability of
successful cheat compared with the party who have not (see Eq.
(\ref{GHZc})).

We also note the scheme can generalize to multipartite case provided
Alice possesses a multipartite entangled state. Suppose she has a
(\textit{N}+1)-qubit entangled state, qubits
$2,3,\cdot\cdot\cdot(\textit{N}+1)$ are sent to \textit{N} users,
respectively. After she confirms that each of the users have
received a qubit, she then operates one of the four local
measurements on qubit 1. After that, the two cbits information was
encoded into the (\textit{N}+1)-qubit entangled state. Later, she
sends her pntonic qubit to one of the rest \textit{N} users. Again,
he who received the photonic qubit will have a higher probability of
successful cheat compared with the rest (\textit{N}-1) users. Only
with the cooperation of all the rest users, one can obtain Alice's
information. In this way, we set up a multipartite secure QDC
procedure.

Next, we will discuss the experimental feasibility of the current
scheme. Among the variety of systems being explored for hardware
implementations for quantum communication, cavity QED system is
favored because of its demonstrated advantage when subjected to
coherent manipulations. The interaction time can be perfectly
controlled by a velocity selector \cite{8}. Relaxation rates of the
system are small and well understood. The strong-coupling conditions
are readily fulfilled \cite{Bertet}. The time constants involved are
long enough to realize all the involved manipulations. Finally, the
quantum systems are separated by centimeterscale distances, thus can
be individually addressed. In our paper, we  assume that the photon
detector is a perfect one and no dark counts. If the finite quantum
efficiency and the dark counts of the detector is taken into
consideration, the fidelity and probability of success will
decrease. For an ordinary photon detector, the dark counts will
reduce the fidelity of the current scheme on the order of
$5\%\sim10\%$. If we make use of Rydberg atoms with principal
quantum numbers 50 and 51, the atomic radiative time can reach
$t_{r} = 3.0\times10^{-2}$s \cite{tr}. Within the current
technology, the quality factor of a cavity can reach $Q =
3.0\times10^{8}$ \cite{8,tr}, thus the effective cavity decay time
can reach $t_{d}=3.0\times10^{-3}$s \cite{9}. Here the coupling
constant $\Omega$ and $g$ can be carefully chosen to reach 10MHz,
and the frequency detune amount can reach 100MHz \cite{14}, the
requirements $\Omega g/\Delta ^2 < < 1$, $\Delta>>\gamma$ and
$\Omega_{k}>>k$ can be satisfied. Using the above mentioned typical
coefficients, we get that the time for the entanglement transfer is
$t_{1} = 1.0\times10^{-4}$s. If we set the time interval for the
detection stage to satisfy $t_{2} = 5.0\times10^{-5}$s, the time
required to complete the protocol is on the order of
$1.0\times10^{-4}$s. In addition, the finite disentanglement time
for an atomic entangled state is $T_{d}=1.6\times10^{-2}$s
\cite{td}. So, the time required to complete the process is much
shorter than the atomic radiative time, the effective cavity decay
time and the finite disentanglement time for an atomic entangled
state. Thus the current scheme might be realizable with the current
cavity QED technology.

In summary, we have investigated a scheme for QDC with GHZ type
entangled state via cavity decay and liner optics devices. The
scheme is probabilistic but secure one. If and only if they
cooperate with each other, they can read Alice's operational
information. Any attempt to get complete information without the
cooperation of the third party cannot be succeed in a deterministic
way. Our scheme combines two distinct advantages: atomic qubit
sevres as stationary bit and photonic qubit as flying bit, thus it
is suitable for long distance quantum communication. In the scheme,
we implemented  all the operation and they were all within current
techniques, thus our suggestion may offer a simple and easy way of
demonstrating secure QDC experimentally.

\begin{acknowledgments}
This work is supported by the Key Program of the Education
Department of Anhui Province (No. 2006kj070A), the Talent Foundation
of Anhui University and the Postgraduate Innovation Research Plan
for from Anhui University entitled to Z.-Y. Xue.
\end{acknowledgments}

\end{document}